# High-precision CTE measurement of hybrid C/SiC composite for cryogenic space telescopes

K. Enya [a,*], N. Yamada [b], T. Imai [c], Y. Tange [c], H. Kaneda [d], H. Katayama [c], M. Kotani [e], K. Maruyama [f], M. Naitoh [c], T. Nakagawa [a], T. Onaka [g], M. Suganuma [c], T. Ozaki [h], M. Kume [i], M. R. Krödel [j]

**Abstract**

This paper presents highly precise measurements of thermal expansion of a "hybrid" carbon-fiber reinforced silicon carbide composite, HB-Cesic® – a trademark of ECM, in the temperature region of ~310−10K. Whilst C/SiC composites have been considered to be promising for the mirrors and other structures of space-borne cryogenic telescopes, the anisotropic thermal expansion has been a potential disadvantage of this material. HB-Cesic® is a newly developed composite using a mixture of different types of chopped, short carbon-fiber, in which one of the important aims of the development was to reduce the anisotropy. The measurements indicate that the anisotropy was much reduced down to 4% as a result of hybridization. The thermal expansion data obtained are presented as functions of temperature using eighth-order polynomials separately for the horizontal (XY-) and vertical (Z-) directions of the fabrication process. The average CTEs and their dispersion (1σ) in the range 293−10K derived from the data for the XY- and Z-directions were $0.805 \pm 0.003 \times 10^{-6}$ K$^{-1}$ and $0.837 \pm 0.001 \times 10^{-6}$ K$^{-1}$, respectively. The absolute accuracy and the reproducibility of the present measurements are suggested to be better than $0.01 \times 10^{-6}$ K$^{-1}$ and $0.001 \times 10^{-6}$ K$^{-1}$, respectively. The residual anisotropy of the thermal expansion was consistent with our previous speculation regarding carbon-fiber, in which the residual anisotropy tended to lie mainly in the horizontal plane.

Keywords    CTE; C/SiC; composite; mirror; telescope; astronomy; cryogenic; infrared

*Corresponding author:

Address: Institute of Space and Astronautical Science, Japan Aerospace Exploration Agency, 3-1-1 Yoshinodai, Chuo-ku, Sagamihara, Kanagawa 252-5210, Japan

E-mail address: enya@ir.isas.jaxa.jp

(a) Institute of Space and Astronautical Science, Japan Aerospace Exploration Agency, 3-1-1 Yoshinodai, Chuo-ku, Sagamihara, Kanagawa 252-5210, Japan

(b) National Metrology Institute of Japan, Advanced Industrial Science and Technology, 1-1-1 Tsukuba chuo-3, Umezono, Tsukuba, Ibaraki 305-8563, Japan.

(c) Earth Observation Research Center, Japan Aerospace Exploration Agency, 2-1-1 Sengen, Tsukuba, Ibaraki 305-8505, Japan

(d) Graduate School of Science, Nagoya University, Chikusa-ku, Nagoya, Aichi 464-8602, Japan

(e) Aerospace Research and Development Directorate, Japan Aerospace Exploration Agency, 6-13-1 Ohsawa, Mitaka, Tokyo 181-0015, Japan

(f) Environmental Test Technology Center, Japan Aerospace Exploration Agency, 2-1-1 Sengen, Tsukuba, Ibaraki 305-8505, Japan

(g) Department of Astronomy, Graduate School of Science, The University of Tokyo, 7-3-1, Bunkyo-ku, Tokyo 113-0033, Japan

(h) Composites Research and Development Co. Ltd., 5-20-40 Kamoi, Midori-ku, Yokohama, Kanagawa 226-0003, Japan

(i) Advanced Technology R&D Center, Mitsubishi Electric Corporation, 1-1-57 Miyashimo, Sagamihara, Kanagawa, 229-1195, Japan

(j) ECM Engineered Ceramic Materials GmbH, D-85452 Moosinning, Germany

1. Introduction

The development of light-weight mirrors is one of the key issues for infrared astronomical space telescope missions, in which telescopes are usually cooled to cryogenic temperature in order to achieve high sensitivity. For space infrared telescopes, material of the silicon carbide (SiC) family has suitable properties, i.e., high ratio of stiffness over mass, high thermal conductivity, and high tolerance against the impact of energetic particles. For instance, the telescopes of AKARI [1] and the Herschel Space Observatory [2], are made of multilayered SiC consisting of a porous SiC core and a chemical vapor deposited SiC coating, and sintered SiC (SiC-100), respectively.

The Space Infrared telescope for Cosmology and Astrophysics (SPICA) is an astronomical satellite mission scheduled to be launched around 2020 [3,4,5]. In the current design, SPICA will have a 3m class telescope cooled to 6K, in which the candidates for the material for both the telescope mirrors and the structure are SiC and carbon-fiber reinforced SiC (C/SiC) composite. One of the advantages of C/SiC over SiC is its high fracture toughness given by the carbon-fiber, while another important advantage is its suitability for machining at the green body phase quite up to the final shape, because the shrinkage of C/SiC is less than 0.5% associated with a very low uncertainty of 0.1%. On the other hand, the anisotropic thermal expansion of C/SiC, which is believed to be due to the carbon-fiber being laid in the horizontal direction (XY-direction) in the manufacturing process, has previously been regarded as a problem. Indeed, the difference in the average coefficients of thermal expansion (CTE), derived by averaging between ambient and cryogenic temperatures, in the XY-direction and vertical direction (Z-direction) of previously manufactured C/SiC is ~20%.

Recent work by ECM in Germany, and Mitsubishi Electric Corporation (MELCO) in Japan [6,7,8] (see also [9]) has led to the development of HB-Cesic®, "hybrid" C/SiC. "Hybrid" indicates that a mixture of different types of chopped, short carbon-fiber is used. As a result of hybridization, some important properties of the material, especially the isotropy and homogeneity, are expected to be improved from previous C/SiC composites. The basic material properties of HB-Cesic® and additional details are given in references [7,8].

For the development of a cryogenic space telescope, it is essential to secure the thermal expansion precisely over a wide temperature range from ambient to cryogenic temperatures, as presented for SiC-100 (Enya et al. 2007; hereafter paper−I) [10]. Especially for the case of C/SiC, a check on the anisotropy of the thermal expansion is also needed. In this paper, we report the results of high precision measurements of the thermal expansion of HB-Cesic® including an estimation of the anisotropy.

## 2. Experiment and result

The instrument set and methods used in this work are basically the same as those used for measuring SiC-100 [10,11]. The instrument set in the Advanced Industrial Science and Technology (AIST) in Japan consists of a cryogenic mechanical Gifford-McMahon cycle refrigerator (V204SC, by Daikin Industries, Ltd.), an interferometer utilizing acousto-optical modulators, a stabilized He-Ne laser system (05STP905, by Melles Griot, Inc.), and a cryostat including a sample holder with optics for interferometric measurements and a temperature sensor (Yamada & Okaji 2000; hereafter paper–II) [11]. Using the same instrument minimizes the systematic errors depending on each instrument enabling precise comparison of the thermal expansion data of different materials.

Fig. 1 shows four samples of HB-Cesic® manufactured and measured in this work. The specification for the size of the samples is $20.00^{+0.05}_{-0.00}$ mm × $20.00^{+0.05}_{-0.00}$ mm × $6.0^{+0.1}_{-0.0}$ mm. The samples were obtained from one blank in which the largest dimension was in the XY-direction in the manufacturing process. The blue and red arrows in the Fig. show the XY- and Z-directions in manufacture, respectively. For each sample, the four 20.00 mm × 6.0 mm surfaces were polished for the measurements, including estimation of the anisotropy.

The obtained data of the thermal expansion are presented in Fig. 2 (a). Each of the data from eight measurements are fit to eighth-order polynomials, $\Delta L/L = \sum_{i=0}^{8} a_i T^i$, and normalized at 293 K as did in paper−I, in which $L$, $\Delta L$, $\alpha_i$, and $T$ are the length of the sample, its expansion, the coefficient of the polynomial, and temperature, respectively. Fig. 2 (b) shows the residual dispersion of the data shown in Fig. 2 (a) from the curves obtained by fitting all the data to an eighth-order polynomial. In Fig. 2, the curves for the data for the XY- and Z-directions clearly separate, and the separation is significantly larger than the dispersion confirmed in a previous measurement for SiC-100 [10]. Therefore, we present the eighth-order polynomials separately for the XY- and Z-directions in Table 1. The average CTEs and their dispersion (1σ) derived from data in the range 293−10K, which is the same temperature region used in paper−I, are $\alpha_{XY} = 0.805 \pm 0.003 \times 10^{-6}$ K$^{-1}$ and $\alpha_Z = 0.837 \pm 0.001 \times 10^{-6}$ K$^{-1}$ for the XY- and Z-directions, respectively. These values of dispersion are of the same order as the value obtained in the measurement for SiC-100, $0.002 \times 10^{-6}$ K$^{-1}$ [10]. In this work, we derived the anisotropy of the average CTE of HB-SiC®, defined by $(\alpha_Z - \alpha_{XY})/\alpha_Z$, to be 4%.

## 3. Discussion and Conclusion

The dispersion in the measurement data is caused by both the intrinsic diversity of the thermal expansion of the samples and the reproducibility of the measurements, and so gives an upper limit for these two sources of dispersion. Even without considering the absolute calibration, such highly reproducible measurements are quite useful for comparing the critical materials used in a single system, e.g., a space telescope, in order to analyze the stress and deformation caused by mismatches in the thermal expansion. Previously, in paper–II, a measurement was carried out on high-purity single crystal of silicon over the temperature range 320−10K, in order to consider the absolute calibration of the instrument, and the result was compared to the recommended value from CODATA [12]. As the result, the upper limit of the deviation from the recommended value was determined to be $\pm 0.01 \times 10^{-6}$ K$^{-1}$. Data in paper–II also give the dispersion due to reproducibility in the measurements to be better than $\pm 0.001 \times 10^{-6}$ K$^{-1}$. The dispersion of the data obtained in this work is close to the value presented in paper–II. Therefore, in this paper, we simply conclude the intrinsic diversity of the thermal expansion of the samples measured in this work is less than $\pm 0.003 \times 10^{-6}$ K$^{-1}$, whilst it is not easy to categorically conclude the existence of intrinsic diversity.

In order to gain an understanding of the reason for the anisotropy of the thermal expansion, we examined the surface of a sample using an optical microscope. Figures 3 (a) and (b) show images obtained from the centers of surfaces (a) and (b) of sample #1 shown in Fig. 1, respectively. The black parts in the figure correspond to carbon-fiber, while the gray and white parts correspond to SiC and Si, respectively [6, 13]. The difference in brightness between Figures 3 (a) and (b) is due to the microscope lighting. As shown in Fig. 3, the features on surfaces (a) and (b) are basically similar, which is the intention in hybridization. On the other hand, it should be noted that there are single carbon-fibers on surface (a) which can be seen as straight narrow black lines. Some of these are highlighted by red rectangles in Fig. 3 (a). Such single fibers are not so clearly found in Fig. 3 (b) as in Fig. 3 (a). This tendency is common for the other areas of surfaces (a) and (b), and for the other samples. One possible interpretation for this feature is that a small fraction of the carbon-fiber still has a tendency to lie in the XY-direction in the manufacturing process even in the hybridized material. This supposition is consistent with the fact that the thermal expansion derived in this work is larger in the Z-direction than the XY-direction, similar to the supposition presented for the anisotropy of previous C/SiC in which the carbon-fiber works to prevent thermal expansion [6]. For surface (b), it may be revealing to carry out imaging of more samples and to analyze the distortion of the carbon-fiber areas statistically in order to estimate the anisotropy quantitatively.

The curves of *ΔL/L* of HB-Cesic® shown in Fig. 2 (a) are flatter in the low temperature region

($\leq$ 100K) than those of SiC-100 [10]. It is considered that this feature is due to the CTE of the carbon-fiber (0.8 to $1.1 \times 10^{-6}$ K$^{-1}$: c.f., $2.6\times 10^{-6}$ K$^{-1}$ and $3.3\times 10^{-6}$ K$^{-1}$ for SiC and Si, respectively) and its temperature dependence [14, 15, 16]. Such very small shrinkage in the low temperature region is a common feature of C/SiC [6, 7, 8], and the data in this paper indicate that this property is inherited by HB-Cesic®. The very small shrinkage is an advantage for the material used for cryogenic space telescopes, for which the test is much easier at the temperature of liquid nitrogen, 77K, than the temperature of liquid helium, 4.2 K.

As presented in paper–I, the measurement has the precision to provide data of the thermal expansion for the design of the SPICA telescope and to perform a significant check on the dispersion of the thermal expansion. If the anisotropy is fully known beforehand, in principle it will be possible, in designing the cryogenic space telescope, to eliminate the effect of anisotropic deformation through cooling and to achieve required wavefront quality of the telescope. On the other hand, such a solution makes the design, fabrication, and tests needed at both ambient and cryogenic temperatures for the telescope much more challenging. These facts can affect the cost, the risk, and delivery time for the development of the telescope. Therefore, it is important that further reductions in the anisotropy of the CTE be made.

Considering the results of this work, further improvement of the material is ongoing. It is potentially fruitful to measure the thermal expansion of candidate materials not only for the SPICA mission but for many other future space telescopes systematically with high precision using the same instrument.

This paper presents highly precise measurements of the thermal expansion of HB-Cesic® in the range ~300−10K. The anisotropy of the average CTE was much reduced, from ~20% to 4% as a result of hybridization of C/SiC. The thermal expansions are presented as functions of temperature using eighth-order polynomials for the XY- and Z-directions separately. The residual anisotropy of the thermal expansion was consistent with our previous speculation considering laid carbon-fiber.

**Acknowledgement**
This work is supported by Japan Aerospace Exploration Agency (JAXA) and AIST. We are grateful to all relating this work in these institutes.

**Reference**

[1] Kaneda, H., Onaka, T., Nakagawa, T., Enya, K., Murakami, H., Yamashiro, R., Ezaki, T., Numao, Y., Sugiyama, Y., "Cryogenic optical performance of the ASTRO-F SiC telescope", Applied Optics, vol. 44, Issue 32, pp.6823-6832 (2005).

[2] Toulemont, Y., Passvogel, T., Pilbratt, G. L.,de Chambure, D., Pierot, D., Castel, D., "all-SiC telescope for HERSCHEL", Proceedings of the SPIE, Volume 5487, pp. 1119-1128 (2004).

[3] Nakagawa, T., SPICA team, "The next-generation space infrared astronomy mission SPICA", Proceedings of the SPIE, Volume 7731, pp. 77310O-77310O-8 (2010).

[4] Kaneda, H., Nakagawa, T., Enya, K., Tange, Y., Imai, T., Katayama, H., Suganuma, M., Naitoh, M., Maruyama, K., Onaka, T., Kiriyama, Y., Mori, T., Takahashi, A., "Optical testing activities for the SPICA telescope", Proceedings of the SPIE, Volume 7731, pp. 77310Z-77310Z-7 (2010)

[5] Onaka, T., Kaneda, H., Enya, K., Nakagawa, T., Murakiami, H., Matsuhara, H., Kataza, H., "Development of large aperture cooled telescopes for the space infrared telescope for cosmology and astrophysics (SPICA) mission", Proceedings of the SPIE, Volume 5962, pp. 448-462

[6] Ozaki, T., Kume, M., Oshima, T., Nakagawa, T., Matsumoto, T., Kaneda, H., Murakami, H., Kataza, K., Enya, K., Yui, Y., Onaka, T., Kroedel, M., "Mechanical and thermal performance of C/SiC composites for SPICA mirror", Proceedings of the SPIE, Volume 5868, pp. 132-141 (2005).

[7] Krödel, M. R., Ozaki, T., "HB-Cesic® composite for space optics and structures", Proc. SPIE 6666, 66660E (2007).

[8] Krödel, M. R., Ozaki, T., Kume, M., Furuya, A., Yui, Y. Y., Imai, H., Katayama, H., Tange, Y., Nakagawa, T., Kaneda, H., "Manufacturing and performance test of an 800mm space optic", Proceedings of the SPIE, Volume 7018, pp. 70180A-70180A-9 (2008).

[9] Kaneda, H., Nakagawa, T., Onaka, T., Enya, K., Makiuti, S., Takaki, J., Haruna, M., Kume, M., Ozaki, T., "Cryogenic optical measurements of 12-segment-bonded carbon-fiber-reinforced silicon carbide composite mirror with support mechanism", Applied Optics, vol. 47, Issue 8, pp.1122-1128 (2008).

[10] Enya, K., Yamada, N., Onaka, T., Nakagawa, T., Kaneda, H., Hirabayashi, M., Toulemont, Y., Castel, D., Kanai, Y., Fujishiro, N., "High-Precision CTE Measurement of

SiC-100 for Cryogenic Space Telescopes", PASP, Volume 119, Issue 855, pp. 583-589 (2007).

[11] Yamada N., Onaji, M., "Development of a low-temperature laser interferometric dilatometer using a cryogenic refrigerator", High Temperatures - High Pressures, 32, pp.199 – 205 (2000)

[12] White, G. K., Minges, M., L. (Eds), CODATA Bull., 59, 13-19 (1985)

[13] Enya, K., Nakagawa, T., Kaneda, H., Onaka, T., Ozaki, T., Kume, M., "Microscopic surface structure of C/SiC composite mirrors for space cryogenic telescopes", Applied Optics IP, vol. 46, Issue 11, pp.2049-2056 (2007)

[14] Catalogue of DIALEAD Carbon Fibers (Mitsubishi Chemical Functional Products, Inc., 2000).

[15] Y. S. Touloukinan, R. K. Kirby, R. E. Taylor, and T. Y. R. Lee, Thermophysical Properties of Matter, the TPRC Data Series, Vol. 13, Thermal Expansion, Nonmetallic Solids (Wiley, 1977) p. 873.

[16] Y. S. Touloukinan, R. K. Kirby, R. E. Taylor, and T. Y. R. Lee, Thermophysical Properties of Matter, the TPRC Data Series, Vol. 13, Thermal Expansion, Nonmetallic Solids (Wiley, 1977) p. 154.

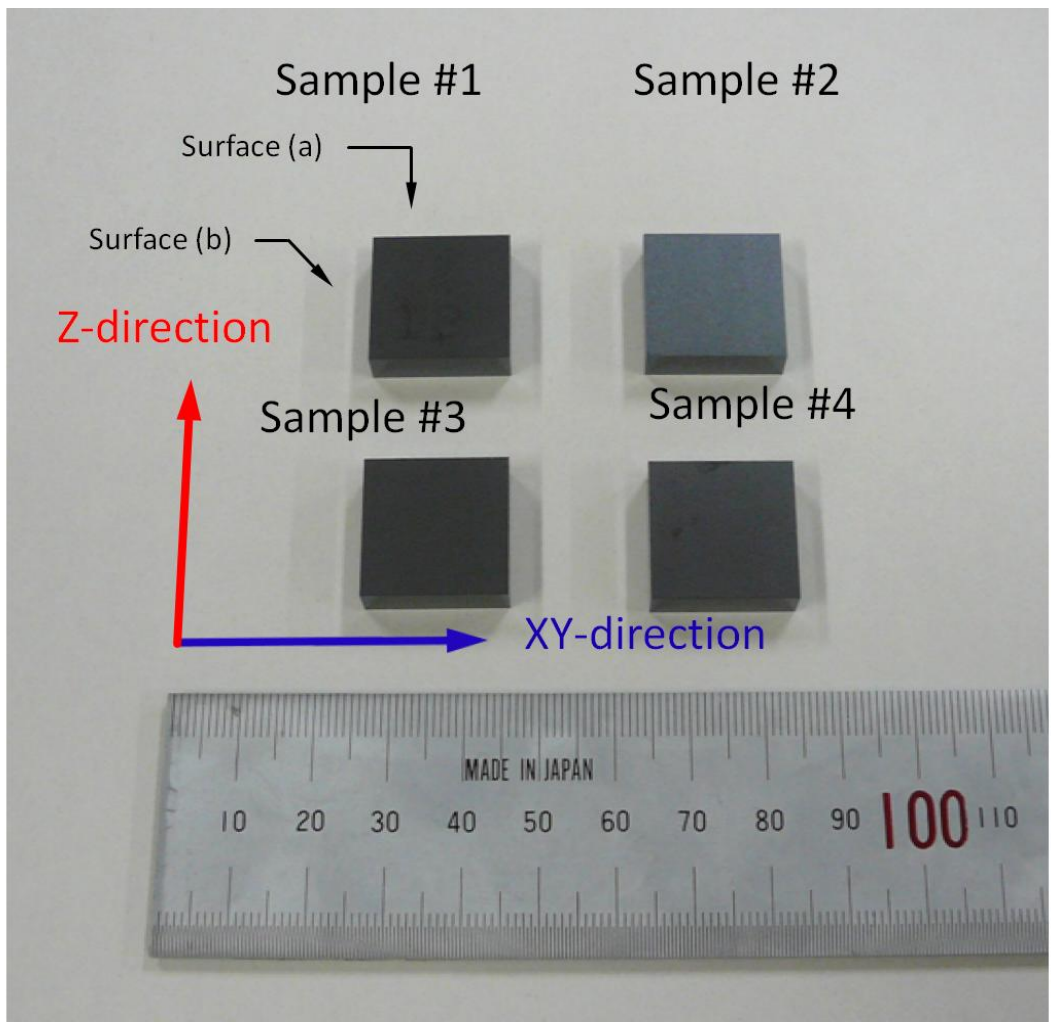

Fig. 1. Samples of HB-Cesic® measured in this work. The dimensions of the samples are 20.00× mm 20.00 mm × 6.0 mm. All of the 20.00 mm × 6.0 mm surfaces are polished. Blue and red arrows shows the XY- and Z-directions. Microscope observations of surfaces (a) and (b) are shown in Fig. 3, respectively.

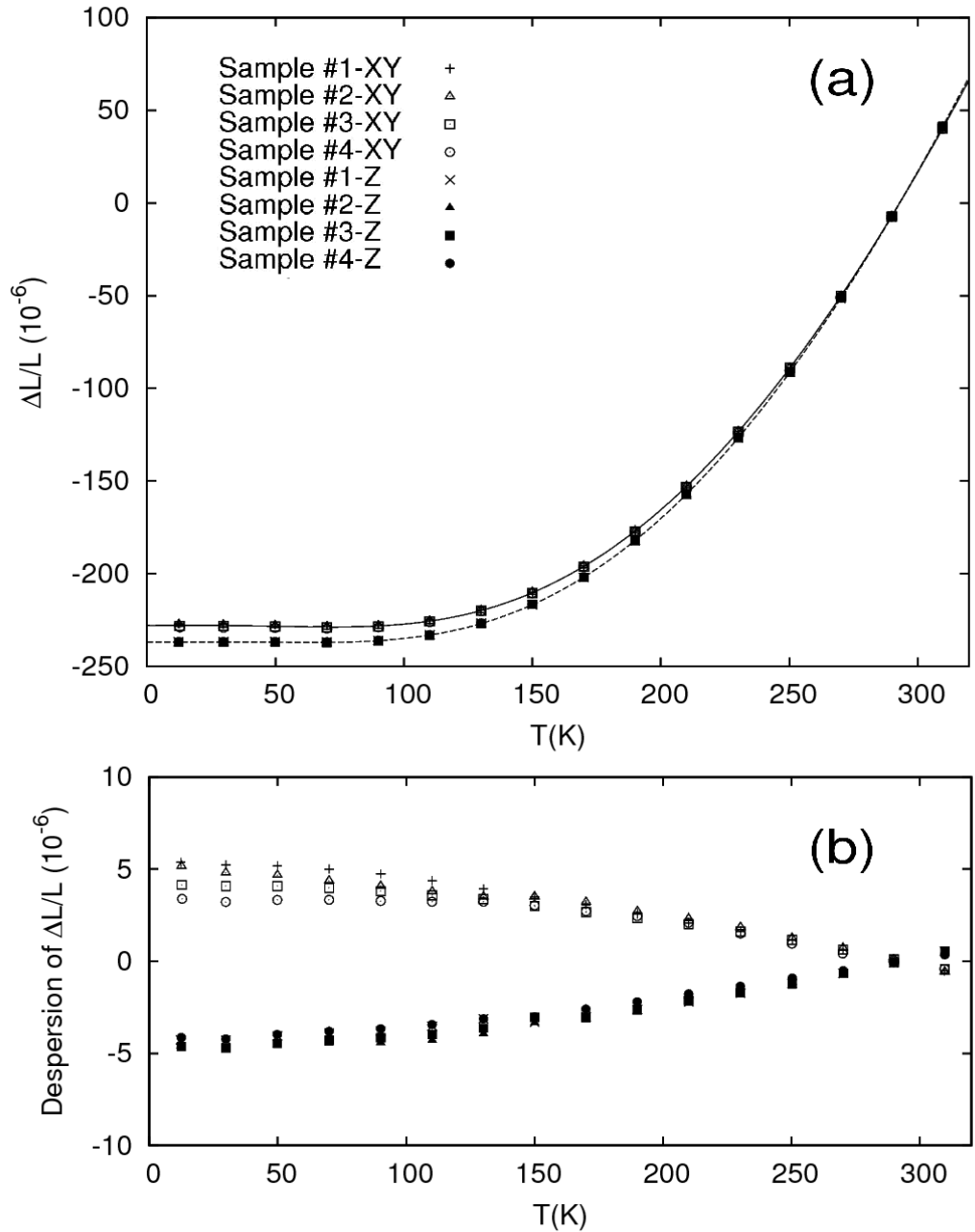

Fig. 2. Data obtained from the measurements. (a): Thermal expansion data for the XY- and Z-directions of samples #1−#4. The solid and dashed lines represent curves fitted to eighth-order polynomials for data of the XY- and Z-directions, the coefficients for which are given in Table 1. (b): Residual dispersion of the thermal expansion data around the fitted curves using all data shown in (a).

Table 1

Coefficients of the eighth-order polynomials

|     | XY-direction | Z-direction |
| --- | --- | --- |
| $a_0$ | $-2.27874 \times 10^{2}$ | $-2.36813 \times 10^{2}$ |
| $a_1$ | $-2.04679 \times 10^{-2}$ | $-2.47792 \times 10^{-2}$ |
| $a_2$ | $+1.86617 \times 10^{-3}$ | $+2.36054 \times 10^{-3}$ |
| $a_3$ | $-6.11583 \times 10^{-5}$ | $-7.07419 \times 10^{-5}$ |
| $a_4$ | $+7.15903 \times 10^{-7}$ | $+8.35768 \times 10^{-7}$ |
| $a_5$ | $-3.40606 \times 10^{-9}$ | $-4.25010 \times 10^{-9}$ |
| $a_6$ | $+8.84326 \times 10^{-12}$ | $+1.21908 \times 10^{-11}$ |
| $a_7$ | $-1.24822 \times 10^{-14}$ | $-1.94229 \times 10^{-14}$ |
| $a_8$ | $+7.54250 \times 10^{-18}$ | $+1.33730 \times 10^{-17}$ |

Note: Coefficients of the eighth-order Polynomial $\Delta L/L = \sum_{i=0}^{8} a_i T^i$ represent the typical thermal expansion of HB-Cesic® in the XY- and Z-directions below ~300 K.

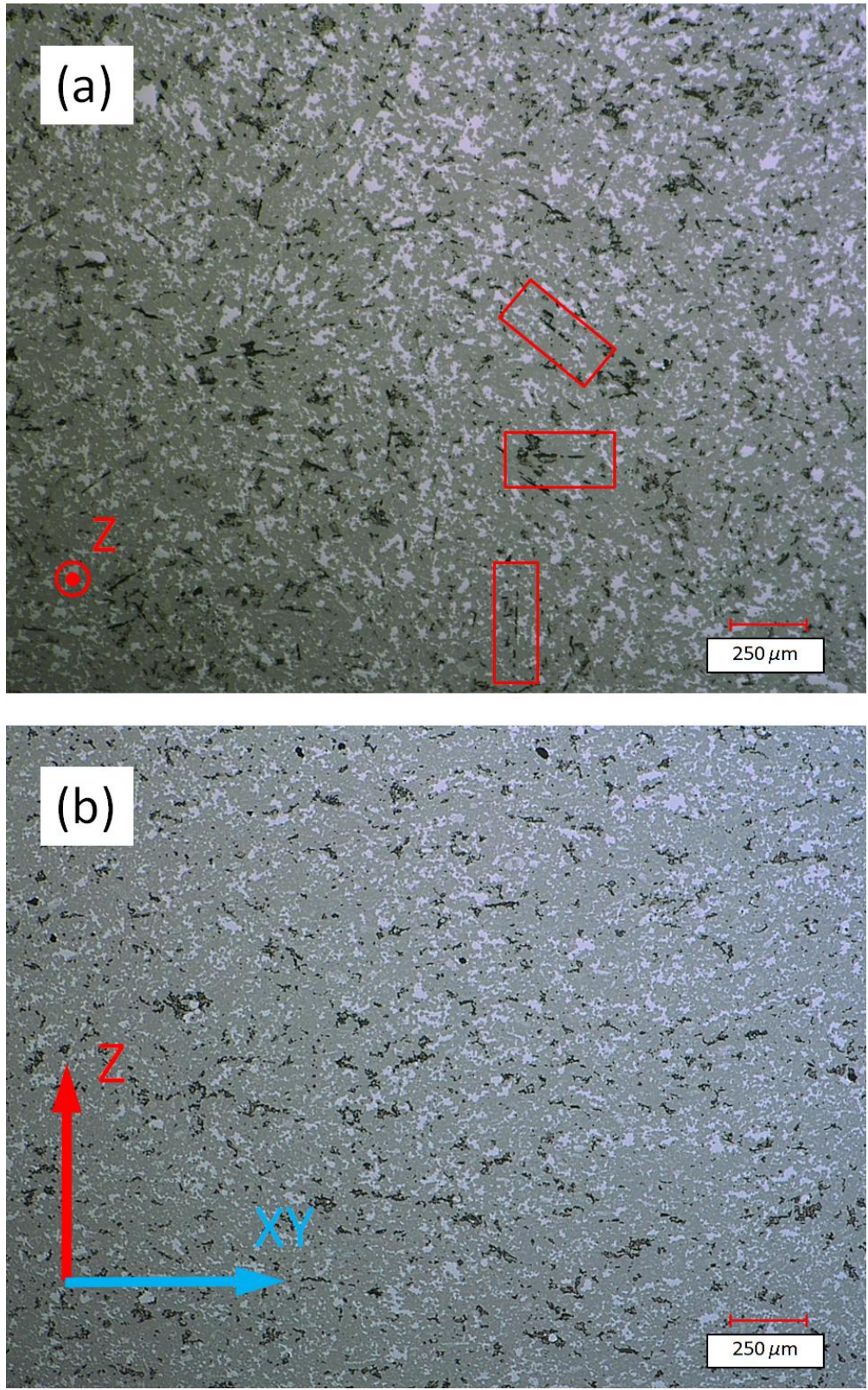

Fig. 3. Surface images of a sample taken with an optical microscope. Panels (a) and (b) show images at the centers of surfaces (a) and (b) of sample #1 shown in Fig. 1, respectively. Blue and red arrows in panel (b) show the XY- and Z-directions. The difference in brightness between (a) and (b) is due to the microscope lighting.